\documentclass{ws-procs9x6-cpt16}
\begin{document}

\newcommand{\refeq}[1]{(\ref{#1})}
\def\etal {{\it et al.}}

\def\fr#1#2{{\textstyle{{#1}\over {#2}}}}

\title{Laser Gyroscopes, Gravity, and the SME}

\author{Nicholas Scaramuzza and Jay D.\ Tasson}

\address{Physics Department, St. Olaf College,
Northfield, MN 55057, USA}

\begin{abstract}
In this contribution to the CPT'16 proceedings,
we illustrate the potential use of ring-laser systems in searching for Lorentz violation 
in the framework of the Standard-Model Extension. 
We present expressions for the Lorentz-violating contribution to 
the ring-laser signal for a sample system
and make sensitivity estimates for the GINGER project.
\end{abstract}

\bodymatter

\section{Basics} 
Experimental discovery of Lorentz violation could provide evidence 
of a unified theory of General Relativity and the Standard Model at the Planck scale.\cite{ks}
The Standard-Model Extension (SME) provides a comprehensive effective field-theory based framework
in which to search for Lorentz violation at accessible energies,\cite{rev}
and a large number of searches have been performed in the SME context.\cite{data}
Ring-laser systems have the potential to provide another interesting test of Lorentz symmetry
in the gravity sector of the SME.\cite{NSJT}
This section briefly summarizes some basic information about ring lasers
and about the gravity sector of the SME
prior to the discussion of our analysis and results in Sec.\ \ref{SME}
and applications in Sec.\ \ref{ginger}.

Ring lasers consist of two light beams traveling in opposite directions along a closed path.  
Effects that break the symmetry of the system 
are encoded in the interference of these beams. 
The largest such effect routinely observed is that of rotating-frame effects on the system 
known as the Sagnac effect, 
which generates a laser beat frequency of
\begin{equation}
\nu_{\rm s} = \frac{4 A \vec \Omega \cdot{\hat{n}}}{P \lambda},
\end{equation}
where $A$ and $P$ are the area and perimeter of the ring respectively,
$\lambda$ is the wavelength of the light,
and $\hat n$ is a unit normal to the ring.
In this way, 
ring lasers are able to measure their own rotation $\Omega$.

A fixed ring laser located on the Earth will experience several effects that alter the beat frequency. 
These include 
the rotation of the Earth, and thus the ring-laser system, 
and frame dragging effects predicted by General Relativity. 
For a planar loop in the equatorial plane,
the gravitomagnetic effects of Earth's angular momentum, 
$\vec{J}_\oplus$, take the following form (in natural units),\cite{lasgyro}
\begin{equation}
\nu_{\rm gm}\approx \frac{4G A J_\oplus}{R^3 P \lambda},
\end{equation}
where $R$ is the radius of the Earth,
and $G$ is Newton's constant.

Since Lorentz violation is known to mimic gravitomagnetic effects,\cite{qbak,lvgmag} 
we consider its impact on ring-laser experiments here. 
Solving the linearized effective Einstein equations for the pure-gravity sector of the minimal SME, 
the following post-newtonian metric is obtained,\cite{qbak}
\begin{eqnarray}
g_{00} &=& -1 + 2U + 3{\bar{s}}^{00} U + {\bar{s}}^{jk} U^{jk} - 4{\bar{s}}^{0j} V^j + ...
\nonumber\\
g_{0j} &=& \bar{s}^{0j}U - \bar{s}^{0k}U^{jk} - \fr{7}{2} \left(1 + \fr{1}{28}\bar{s}^{00}\right) V^j+... 
\nonumber \\
g_{jk} &=& \delta ^{jk} + (2 - \bar{s}^{00})\delta ^{jk}U+\left(\bar{s}^{lm}\delta ^{jk} - \bar{s}^{jl}\delta ^{mk}\right) U^{lm} +...,
\label{metric}
\end{eqnarray}
which forms the starting point for the current analysis. 
Here $\bar{s}^{\mu \nu}$ is a coefficient for Lorentz violation in the minimal gravity sector.
The standard newtonian potential is given by $U$,
and the remaining potentials 
are defined as $U^{j k}=G \int{d^3 x^\prime \rho (\vec{x}^\prime,t)R^j R^k/R^3}$ and $V^j=G \int d^3 x^\prime \rho (\vec{x}^\prime ,t) v^j(\vec{x}^\prime ,t)/R$.\cite{qbak}
Though it lies beyond our present scope,
we note in passing that a similar analysis for the matter sector
may also be of interest.\cite{lvgap}

\section{Sample system}
\label{SME}
Here we consider Lorentz-violating effects stemming from Eq.\ (\ref{metric}) 
that generate signals in ring-laser experiments in the absence of Sagnac or gravitomagnetic effects, 
as opposed to those that can arise as a perturbation on these effects. 
While it is clear that perturbations on the conventional effects also exist, 
resulting signals are typically suppressed. 
Working to leading order in Lorentz violation, 
we derive the beat frequency in analogy with gravitomagnetism 
by calculating the time of flight difference between the two beams 
due to $\bar{s}^{tj}$
via an integral along the light-like world lines of Eq.\ (\ref{metric}).

In this proceedings contribution,
we consider a sample ring-laser system
as an illustration of the results that arise.
For the sample system,
we consider an approximately rectangular loop in the equatorial plane
for which two sides are radial and two sides are azimuthal.
For the radial legs of the path, 
which provide the signal of interest, 
the world line takes the following form:
\begin{equation}
0={d\tau}^2 = dt^2 (1-2 U) + 2 dt dR g_{0j} \hat R^j - dR^2 (1+2 U). 
\end{equation}
Solving for the beat frequency under these conditions
yields
\begin{equation}
\nu_{\rm LV}= \frac{4gA}{ P \lambda} (-\bar{s}^{TX}\sin\phi +\bar{s}^{TY}\cos\phi),
\label{lvnu}
\end{equation}
where $g$ is the newtonian gravitational field
and capital indices $T,X,Y$ denote components in the standard Sun-centered frame.\cite{qbak}
Note that on the surface of the Earth, $\phi = \omega_\oplus t$.
Hence the observable beat frequency varies sidereally
in the presence of Lorentz violation. 
Using the previously described method, 
it is possible to generate the resulting dominant contributions to the beat frequency 
for a ring laser in any orientation.\cite{NSJT}

\section{The GINGER project}
\label{ginger}
The GINGER (Gyroscopes IN GEneral Relativity) experiment will consist of several perpendicular ring lasers 
and is currently being designed with the sensitivity to measure the gravitomagnetic effects 
of General Relativity.\cite{lasgyro}
The experiment expects to obtain sensitivity to the angular velocity of the Earth 
via the Sagnac effect beyond the part in  $10^{9}$ level.\cite{lasgyro} 
Using this experimental sensitivity
and the sample system above,
it is possible to generate a crude estimate of the sensitivity to Lorentz violation 
that might be achieved by the GINGER project. 
We predict that sensitivities to $\bar{s}^{TJ}$ better than parts in $10^6$ are possible.
With the exception of astrophysical tests,\cite{astro}
which perhaps involve additional assumptions,
sensitivities at this level would be competitive with the current state of the art
measurements.\cite{data,hees}

\end{document}